\documentclass[conference]{IEEEtran}
\IEEEoverridecommandlockouts
\usepackage{cite}
\usepackage{amsmath,amssymb,amsfonts}
\usepackage{algorithmic}
\usepackage{graphicx}
\usepackage{textcomp}
\usepackage{xcolor}
\usepackage{balance}
\usepackage[hyphens]{url}
\usepackage{hyperref}
\usepackage[hyphenbreaks]{breakurl}

\def\BibTeX{{\rm B\kern-.05em{\sc i\kern-.025em b}\kern-.08em
    T\kern-.1667em\lower.7ex\hbox{E}\kern-.125emX}}

\begin{document}
\title{``Is My Mic On?''\\Preparing SE Students for Collaborative Remote Work and Hybrid Team Communication}

\author{\IEEEauthorblockN{Makayla Moster\IEEEauthorrefmark{1}, Denae Ford\IEEEauthorrefmark{2}, and
Paige Rodeghero\IEEEauthorrefmark{1}}
\IEEEauthorblockA{\IEEEauthorrefmark{1}School of Computing, Clemson University\\
Clemson, South Carolina 29631 \\
Email: \{mmoster, prodegh\}@clemson.edu}
\IEEEauthorblockA{\IEEEauthorrefmark{2} Microsoft Research \\
Redmond, WA 98052\\
Email: denae@microsoft.com}}

% make the title area
\maketitle

\begin{abstract}
Communication is essential for the success of student and professional software engineering (SE) team development projects. The projects delivered by SE courses provide valuable learning experiences for students because they teach industry-required skills such as teamwork, communication, and scheduling. Professional SE teams have adopted communication software such as Slack, Miro, Microsoft Teams, and GitHub Discussions to share files and convey information between team members. Likewise, they have distributed software development tools such as Visual Studio CodeSpaces and Jira to support productivity.  In contrast, within academia, students have focused on having face-to-face meetings for team communication and communication tools for file sharing. Due to the COVID-19 pandemic, universities have been forced to switch to an online or hybrid modality abruptly, thus compelling SE students to quickly adopt communication software. This paper proposes a study on the use of communication software in industry to prepare students for remote software development positions after graduation.
\end{abstract}

\section{Introduction}
Traditionally, students taking software engineering (SE) courses complete a team project over the semester to apply knowledge learned and to help prepare for future industry roles~\cite{dzvonyar2018team}. These team projects are assigned to teach students industry-required skills such as familiarity with codebases, teamwork, communication,  scheduling, etc~\cite{dorodchi2019teaching}. Unfortunately and recently, the novel coronavirus disease 2019 (COVID-19) pandemic has forced many academic institutions to transition to fully online or hybrid model courses. The pandemic has made it impossible for many students to meet face-to-face to work on their team projects.  This means that student teams must adopt an online communication tool such as the university’s provided learning management system (LMS), GroupMe, Slack, or Microsoft Teams to adapt their collaboration to an online modality. 

Online versions of SE courses are not a novelty of the COVID-19 pandemic. Many universities offered online courses and traditional face-to-face classes before the pandemic started, which normally had a higher enrollment of students from non-traditional backgrounds~\cite{dumford2018online}. The advantages of learning in an online modality include greater accessibility, more flexibility and reduced commute time and carbon footprint~\cite{dumford2018online, serrano2019technology}. Disadvantages of online-only learning include isolation, weaker engagement, and technology interruptions~\cite{dumford2018online, serrano2019technology, ragusa2018sense}. Although there are disadvantages, in 2013 over 5.5 million students were enrolled in at least one online course and almost 70\% of U.S. universities reported online classes as part of their future strategies~\cite{seaman2018grade, allen2016online}. %Though there are disadvantages to online learning, there were over 5.5 million students enrolled in at least one online course in 2013 and almost 70\% of universities in the United States reported online classes as part of their long-term strategies~\cite{seaman2018grade, allen2016online}. 

Unfortunately, due to the COVID-19 pandemic, these disadvantages are exacerbated in students caused by the increased stress, isolation, and uncertainty surrounding everyday life. Solely online learning requires stable internet access and technological devices. Those who lack internet access or rely on campus technology or the internet are at a disadvantage from the unexpected transition to online learning~\cite{adedoyin2020covid}. Online students are also more likely to have a decrease in classroom productivity and learning engagement due to the higher risk of being interrupted by family members or pets~\cite{adedoyin2020covid}.
%Online students are also more likely to have interruptions from pets or family members which take their attention away from class, and decreases their overall productivity and learning engagement~\cite{adedoyin2020covid}.
%Online students are also more likely to have a decrease in classroom productivity and learning engagement due to the higher risk of being interrupted by family members or pets~\cite{adedoyin2020covid}.

%On the other hand, there also are similar drawbacks to fully remote development work in industry. Social isolation and blurred boundaries between home and work are a couple of the most prominent drawbacks during normal, pandemic-free remote work. 

The COVID-19 pandemic has impacted students and software developers' lives by forcing them to abruptly switch from their office to a fully remote workflow. This switch caused developers to be more reliant on their communication software to share ideas and files between team members~\cite{rodeghero2020empowering}. Ford \textit{et al.} conducted a survey of 2,265 software engineer respondents during the coronavirus pandemic to identify the challenges that remote software developers face~\cite{ford2020tale}. They found communication between team members takes twice as long due to increased response times and developers feel pressured to be responsive to work messages at all times~\cite{ford2020tale}.

%Common challenges involved communication between teammates taking twice as long due to increased response times and the pressure to be highly responsive to work messages at all times ~\cite{ford2020tale}.

% Moved the second paragraph down here, I think it makes more sense here than it did before.

%To prepare students for future industry roles, SE instructors incorporate a semester-long development project into their courses.

Before the COVID-19 pandemic, SE instructors typically incorporated a semester-long development project to help prepare their students for future industry roles~\cite{ardis2014software}. Traditionally, these team SE projects help students learn marketable and soft skills, such as version control and communication, through the simulation of a professional SE development team. Communication is a necessary skill for graduates to have for software development positions in industry~\cite{shaw2005deciding, radermacher2014investigating, 10.1145/3194779.3194780}.  For example, global and distributed software development have become increasingly popular in recent years. These teams have developers across the world collaborating on projects. Unfortunately, this distribution of team members can cause additional challenges in collaboration and communication~\cite{stray2019slack, li2016teaching}. Companies rely on software such as Skype, Slack or Microsoft Teams to convey information between team members~\cite{stray2019slack, chatterjee2020software}. In a previous study, one SE course simulated a global development team using strict timezones for team members and communication software~\cite{li2016teaching}. Students could use whichever software they deemed appropriate, with the majority choosing Facebook and Skype for their chatting features~\cite{li2016teaching}. Communication applications make it easier for any team, regardless of location or occupation, to collaborate. %However, when team members have inadequate communication skills, virtual team communication can cause conflict~\cite{layng2016virtual}.

%The overall goal of this work is to identify how the use of communication tools during online or hybrid SE course group projects is preparing students for jobs post-graduation. 

In this paper, we present a proposed evaluation of various communication tools SE student teams use, how productive the teams are, how effective their communication is, and their feelings toward using software instead of face-to-face meetings. We also present a plan to explore the tools currently being used in industry to compare and contrast with the ones being used in academia.  We note that team projects in software engineering courses aim to help students learn and improve the skills necessary for their future roles in industry. Therefore, as the future of work has shifted, in some ways overnight, we need to understand the new industry standards better to prepare students for their careers. The future of the software development industry after the COVID-19 pandemic is uncertain, but by preparing students for remote development roles now, they will be more flexible entering the workforce.

\section{The Problem}

With the sudden introduction to online and hybrid courses due to the COVID-19 pandemic, software engineering courses have to shift their classrooms and material online. Still, instructors continue to assign team projects due to their valuable learning experiences. The problem we aim to address is that due to the online and hybrid modality of courses, SE students must find new tools for communication between team members.  %However, because of the nature of SE courses preparing students for industry. 
Unfortunately, we do not know what tools are currently being used by industry.  Therefore, we must explore tools used for remote communication and work in the industry and apply them to the classroom environment. During team projects, students' communication skills are essential for success in preparing for industry roles, even more so now, due to the industry uncertainty caused by COVID-19. 

Radermacher \textit{et al.} and Anewalt \textit{et al.} identified communication as an area that recent graduates are deficient in their early career as software developers~\cite{radermacher2014investigating, anewalt2017curriculum}.  Communication is a key skill for graduates to have not just for conveying ideas with their future colleagues but also for conversing with future customers. Customer software requirements are often vague and unhelpful, but by acquiring the skills to communicate effectively through coursework, students will be better prepared to elicit the necessary information from customers~\cite{saiedian2000requirements}. Previous studies found that the quality that developers communicated with stakeholders influenced overall customer satisfaction with the developed product~\cite{jolak2020software, jarboe1996procedures, kortum2017don}. Anewalt \textit{et al.} surveyed 113 randomly-selected SE alumni from the University of Mary Washington and found that they preferred incorporating additional communication courses in the university's computer science degree plan, based on their early-career communication experiences, to increase students' oral communication skills~\cite{anewalt2017curriculum}. 
%Their team surveyed 113 randomly-selected SE alumni from the University of Mary Washington and found that based on their early-career experiences, they preferred incorporating additional communication courses to the CS degree plan to increase students' oral communication skills~\cite{anewalt2017curriculum}. %Anewalt \textit{et al.} surveyed 113 randomly-selected SE alumni from the University of Mary Washington and found that they would prefer incorporating additional courses to the computer science degree plan to increase current students’ development in oral communication due to their early career experiences~\cite{anewalt2017curriculum}.

%Anewalt \textit{et al.} surveyed 113 randomly-selected SE alumni from the University of Mary Washington and found that based on their early-career experiences, they prefered incorporating additional communication courses to the computer science degree plan to increase students' oral communication skills~\cite{anewalt2017curriculum}.

By simulating an industry-level project in the classroom, Iacob \textit{et al.} found in their study that SE students perceived team communication as a challenge during the beginning of a team project and that the students preferred to have face-to-face meetings over remote communication via industry-accepted communication platforms such as Slack~\cite{iacob2019exploring}. Unfortunately, due to the COVID-19 pandemic, the option for face-to-face team meetings is unfeasible, and students are required to transition to online software for communication with their teammates. The ability to effectively utilize online tools for communication is crucial to the future of the software development industry as remote work becomes more desirable for employees' safety and company budgets. This proposal seeks to identify the communication tools utilized by SE student and industry teams and communication effectiveness, productivity, and feelings toward using software to communicate to further prepare SE students for future remote development roles.

\section{Background and Related Work}
In this section, we discuss the background and related work on online education, industry team project communication, software engineering team project communication, and remote work communication challenges.
\subsection{Online Education}

Online education has become increasingly popular with universities in recent years, with one 2016 study finding that online or distance education enrollments had increased for the past fourteen years. It found that about 30\% of all students enrolled in at least one online course~\cite{allen2016online}. A 2018 survey conducted by the U.S. Department of Education found that over 6 million students, or 6\% of all students, take at least one online course~\cite{seaman2018grade}. The same study found that the total number of solely on-campus student enrollments dropped by 6.4\%, or over 1 million students, between 2012 and 2016~\cite{seaman2018grade}. 

There are two standard models for online education, hybrid/blended or fully online. The hybrid (or blended) model is where part of the students’ learning material is delivered online, and the other part is delivered in-person. The fully-online model does not require students to attend a physical classroom due to all learning materials being online~\cite{blayone2017democratizing}.

With the increased demand for online education in the era of portable smart devices, students are more likely to use mobile versions of LMS over desktop counterparts~\cite{dumford2018online}. Al-Emran \textit{et al.} found in their study 99\% of students have mobile devices and 41.5\% of them use mobile devices for web browsing and accessing email~\cite{al2016investigating}. Previous research suggests that using mobile devices for online learning results in positive outcomes~\cite{dumford2018online}. However, there are drawbacks to solely online education, such as isolation, reduced faculty-student interaction, and technological interruptions~\cite{dumford2018online}.

\subsection{Team Communication in Industry}

There is a rich body of literature on communication within software development teams. Communication is an important process within the development of software. Without communication between team members, many projects would fail immediately. Ineffective communication between project team members is much harder to fix than improving a team member's technical skills~\cite{deFranco2017review}.

Face-to-face interactions are an important communication method between developers since it allows for immediate feedback and discussion~\cite{storey2017how}. However, there are different tools that development teams use for communication between projects. Due to the increase in global or distributed programming, most development teams have to adopt communication and organization tools between members~\cite{storey2017how}. Some more popular team communication platforms include Slack, Flowdock, and Hipchat~\cite{anders2016team}. These different software platforms support collaborative chat threads organized into distinct categories for different project aspects. They have multiple social features, such as instant messaging, video calls, separate channel topics, and the social connectivity that social networking platforms provide~\cite{anders2016team}. Storey  \textit{et al.} found in their study that developers prefer to replace face-to-face discussions with private chats when teams are distributed~\cite{storey2017how}.

Team communication platforms are changing how industry developers communicate with one another. One news outlet has called Slack an ``email killer'' due to the reduced need to send emails between team members~\cite{anders2016team}. Anders \textit{et al.} found in their study that adopting a team communication platform can have meaningful changes in individual and team collaboration and communication~\cite{anders2016team}. Though there are drawbacks to virtual communication. Storey \textit{et al.} found developers felt distracted by the constant stream of notifications \cite{storey2017how}. There was also a rise in miscommunication and information fragmentation between communication channels and an overabundance of information for developers~\cite{storey2017how}.

\subsection{Student Team Project Communication}

SE team projects are valuable learning experiences for students who have no prior practical experience. Since these projects are important to shape inexperienced developers, there is a large body of literature on the topic. Team projects teach students different skills they will utilize in industry, such as teamwork, communication, and scheduling. To make these projects as industry-related as possible, some universities team up with industry partners to come up with real problems for students to solve. However, most universities do not have the resources to include industry projects in their classrooms but allow students to use industry-level communication tools. 

Seppala \textit{et al.} conducted a study to determine what tools student SE teams were choosing to use for communication and sharing information between members~\cite{seppala2016communication}. They found that face-to-face communication between members was the most preferred method. However, students chose to use other methods such as WhatsApp, Trello, GitHub Issues, Skype, social media, Piazza, and Google Hangouts~\cite{seppala2016communication}. 

Another study chose to simulate global SE projects from within the classroom, allowing student teams to choose their communication platforms and different simulated timezones for each team member to work in~\cite{li2016teaching}. During their project, they found that 21\% of their students used Skype to chat between team members and 61\% used Facebook to chat~\cite{li2016teaching}.

These two examples show that communication is essential during team projects, and that students can identify tools that they need to use to communicate effectively with their teammates. Our study seeks to determine if the tools students choose to use for team communication help prepare them for post-graduation remote software development roles.

\subsection{Remote Communication Challenges}
Before the COVID-19 pandemic, business or team meetings were not limited to physical interactions. Companies increasingly held virtual meetings to include their stakeholders or geographically-distributed team members for discussions~\cite{layng2016virtual}. The advantages of remote communication include reduced travel cost, teammates' ability to have regular contact through software, and increased flexibility~\cite{layng2016virtual, handke2018medium, ford2019remotework}. However, although remote communication has become increasingly popular, some challenges arise.

Layng \textit{et al.} found in their 15-year survey of the literature that virtual communication generates many interpersonal challenges in teams~\cite{layng2016virtual}. These challenges involve communication frequency and effectiveness, teammate trust, knowledge transfer, and coordination~\cite{jimenez2017working, layng2016virtual}.

Communication is fundamental for building relationships and trust between individuals~\cite{jarvenpaa1999communication}. Members of teams located in the same office space are more likely to have social interactions outside of the office; therefore building relationships and trust in each other. However, virtual teammates cannot join in in-person social activities and instead rely on virtual communication methods to build trust with their teammates. Face to face communication allows for persons to provide nonverbal cues that build trust between the team, which are unavailable in text-based virtual communication channels~\cite{jimenez2017working}. Previous research suggests that remote teams should create trust by engaging in virtual water cooler conversations and utilizing video conferencing methods at least two weeks before the project start date~\cite{layng2016virtual}. 

%Previous research suggests that for successful projects, virtual teams should take part in virtual water cooler conversations and use video conferencing methods to create trust between team members two weeks before beginning a project to ensure success~\cite{layng2016virtual}.

Communication for virtual teams is often less frequent than that of in-person teams~\cite{layng2016virtual}. Continuous communication between teammates is required to resolve conflicts and avoid teammate confrontations or miscommunication~\cite{rutowske2002}. Without clear communication guidelines or schedules, team members become silent, causing disastrous team conditions such as frustration or miscommunication~\cite{layng2016virtual}. These conditions negatively affect teammate trust and coordination and can cause project failure if not remedied quickly. Multiple studies found that businesses were using multiple communication channels for their virtual teams such as email, instant messaging, and video conferencing to reduce this miscommunication~\cite{layng2016virtual}. These channels should also be chosen based on the team's tasks to reduce miscommunication and confusion further.

Overall, frequent communication and teammate trust are essential for projects with virtual teams to be successful. Key challenges in remote communication involve building trust between teammates, maintaining frequent communication, coordinating tasks, and communicating effectively. Previous studies have suggested setting communication ground rules to mitigate these issues and start using video-conferencing mechanisms to build trust between team members~\cite{rutowske2002, layng2016virtual}.

\section{Proposed Study Design}

In this section, we describe our research questions and proposed methodology for studying remote collaboration tools.

\subsection{Research Questions}

The objective of this proposal is to evaluate the communication tools used by industry and student teams, to determine how productive the teams are, the effectiveness of their communication, and personal perceptions toward using software for communication in place of face-to-face meetings. The main question we want to begin to answer is \textit{are online and hybrid classes preparing students for future remote software developer roles?} %We will compare and contrast the communication tools used in industry with the tools found in academia for recommendations to improve overall team communication tool usage. 

    %The objective of this proposal is to evaluate the communication tools used by industry and SE student teams, how productive the teams are, as well as how effective their communication is and their feelings toward using software in place of face-to-face meetings. 
    
Therefore, we ask the following research questions:

\begin{description}
\item[$RQ_{1}$] What communication tools are currently being used by industry?
\item[$RQ_{2}$] What communication tools are used by student teams?
\item[$RQ_{3}$] What do teams define as effective communication while working remotely?
\item[$RQ_{4}$] How can we get student developer communication tools to be more reflective of industrial development while still staying learning-centric?

% What are the general perceptions of commonly used communication tools?

\end{description}

The rationale behind $RQ_1$ is to determine the current tools being used by industry development teams. Once we know the tools used in industry, we can compare and contrast them with the tools found in $RQ_2$ to determine which ones have overlapping utilization in academia and industry. 

$RQ_3$ aims to determine if SE student teams can communicate effectively when using online communication tools. This question will explore how well teams communicate virtually, leading to $RQ_4$, which will investigate how academic uses of communication software differ from uses in industry. These questions aim to determine if teams communicate effectively, their feelings related to team project virtual communication, and their thoughts on face-to-face meetings instead of using communication software. Additionally, these questions will help the community better understand the differences between remote work and remote education so that we are able to improve the education of our students for industry careers. 

By exploring academic and industry remote communication methods, we can gain greater knowledge about remote communication utilization's strengths and weaknesses to further our understanding of classroom and industry standards. By comparing industry and academic software utilization, we will make recommendations for future academic courses to implement remote communication to prepare students for future remote developer roles. We will also make feature recommendations to software companies based on the requests gathered from current developers and students.

%We will also be able to make software suggestions to industry based on input from developers

%By exploring the different remote communication methods of industry and academia, we can investigate the differences and similarities between them to gain greater knowledge about what tools and communication standards are being used in industry. Then we can convey our findings to SE classrooms to better prepare our students for their future careers.

%We want to determine if the team members like they can communicate effectively with their software, if they like communicating virtually, and if they would rather have face-to-face meetings.

\subsection{Methodology}

This subsection describes the proposed methodology to answer the research questions from the previous subsection. We plan to recruit two groups of participants, one made up of industry developers and the other undergraduate students in academia, through social media advertisements. 

%moved here for better clarity

The information gathered from student and industry developers will be divided across multiple phases. First, we will gather preliminary information from surveys, and then we will conduct semi-structured interviews to get a more in-depth discussion about current collaboration practices in academia and industry. All surveys will be pilot-tested before distribution online. The advertisements for both student and industry developers will include the survey to gain knowledge about their communication software. The survey will contain instructions at the end to contact members of the research team to schedule an optional semi-structured interview.

The advertisements for undergraduate college students will include a survey to complete student status questions and their experience with teamwork and online learning. The surveys will include a mix of multiple-choice and short answer questions. The students' survey answers will help us answer $RQ_2$ and $RQ_3$ and $RQ_4$ from the student perspective. We will ask preliminary questions such as class level, university attended, programming experience, course delivery method, and major. To ensure that our student respondents have basic programming knowledge, we will include a simple snippet of code for the students to analyze and answer questions after the preliminary student-status questions. The students will then answer questions related to their remote education group project experiences, such as what communication tools were used by their team, their personal feelings about the communication tools, their ability to communicate with their team members effectively, and their thoughts on the team's overall communication effectiveness. At the end of the survey, students will be instructed to contact a member of the research team to schedule an optional interview. The interviewer will ask questions such as how the student teams use collaboration tools, their feelings on communication with teammates, and any ideas to improve the communication and collaboration software for future student teams. 

% The information gathered from industry developers will be divided across multiple phases. First, we will gather preliminary information from surveys, and then we will conduct semi-structured interviews to get a more in-depth discussion about current industry collaboration practices. The advertisements for industry developers will include the survey to gain knowledge about their communication software. They will contain instructions to contact members of the research team to schedule an interview. 

For the industry developer surveys, there will be a mix of multiple-choice and short answer questions about the developer's current position, collaboration tools, feelings pertaining to the use of collaboration tools, effectiveness level of the tools for development roles, and overall thoughts on how well the tools are used in team communication. The developers' survey answers will help us determine the current tools used by industry to answer $RQ_1$. The developer will then schedule an optional appointment with a member of the research team for an interview to delve further into their thoughts and feelings on the collaboration tools used in their development team, which will help us answer $RQ_3$ and $RQ_4$ from the developers' perspective. The interviewer will ask questions such as how the developers use the tools for collaboration if there are different tools for different tasks, their feelings on how well they communicate with their team, and any improvements they wish they could have within their communication software. 

\subsection{Data Collection}

Our data collection will include the participants' demographics, programming experience, occupation and education level, and information about their virtual team communication methods. Our interviews with students and industry developers will be held via video conferencing software or phone call and recorded for referral, coding, and transcription. In addition, we are exploring the idea of data mining communication tools, such as Slack, to understand communication within teams. 

\subsection{Participants}

By posting our advertisements on social media platforms, we aim to gather information from a diverse group of students and developers from a wide variety of colleges and companies.

We plan to recruit undergraduate SE student participants by posting survey advertisements on popular social media outlets such as Reddit and Twitter to acquire a sample from multiple colleges. To ensure those who respond to our survey have a technological background, we will include code snippets for the respondents to analyze and answer questions. 

We will also recruit developers from industry by posting our advertisement on social media such as LinkedIn, Twitter, and Reddit. To ensure those who respond are actual developers, we will also have a code snippet question included in the survey. To recruit developers for the interview portion of our study, we will include instructions within the survey to contact a member of the research team to set up an interview appointment.

\subsection{Analysis}
%frequency of communication methods used by each group.
After completing all surveys and interviews, we will compile and conduct a thematic analysis using open coding on all of the responses to determine common themes within the data. Coding the responses will give an overview of how frequently specific tools are used for certain collaborative activities such as sharing files, chatting with team members, or generating ideas. The codes will also allow our team to determine which improvement requests are the most popular with developers and students. Then we will complete a thorough comparison of the methodologies used in academia and industry to determine the overlap in software choices and uses and improvement requests. Based on this comparison, we will better understand the similarities and differences in how students and industry developers use software to collaborate.

\section{Limitations and Threats to Validity}

Similar to all research, there are limitations to how this study was designed. First, due to distributing our survey on social media platforms, responses will be limited to students and developers who have access to social media platforms such as LinkedIn, Twitter, Reddit, Facebook, etc. Second, student surveys and interviews may be completed during the course project or after their course project is completed. Therefore, some students may be answering questions post-course and may have forgotten their feelings toward the communication software. Other students may only have early impressions of the communication software and have not formed solid opinions due to lack of usage. Third, since respondents are self-selecting to answer our survey questions, non-response and volunteer bias may also skew our results. Furthermore, respondents could cause unintentional bias due to their previous work or university-related experiences.

\section{Conclusion}

This idea paper presented a proposed evaluation of various communication tools used by industry and student teams and team productivity, communication effectiveness, and feelings toward using communication software instead of in-person meetings. We also presented a plan to investigate further industry developers' communication tools to compare with the tools being used in academia. With our findings, we will make recommendations to both academia and industry usage of communication tools to further improve remote team communication effectiveness. As a community, we will also improve communication tools based on our interviews with industry and student developers. Due to the COVID-19 pandemic, the future of work has shifted faster than predicted, and to better prepare students for future remote industry roles, we need a greater understanding of industry standards. Our findings will shed new light on industry developers' communication tools, and our comparison with the current tools used by students will lead to tool recommendations for academic adoption to help instructors better prepare their students for future industry careers. Therefore, if we prepare students for remote roles that are becoming increasingly popular (such as positions at Stack Exchange, GitLab , GitHub, etc.), they will have a significant advantage in finding software engineering careers of the future. We would like to receive as much feedback as possible from the software engineering research community on the proposed research. We welcome researchers, teachers, and software engineers looking to collaborate on this work.

\balance
% references section
\bibliographystyle{IEEEtran}

\bibliography{biblio}

% Generated by IEEEtran.bst, version: 1.14 (2015/08/26)
\begin{thebibliography}{10}
\providecommand{\url}[1]{#1}
\csname url@samestyle\endcsname
\providecommand{\newblock}{\relax}
\providecommand{\bibinfo}[2]{#2}
\providecommand{\BIBentrySTDinterwordspacing}{\spaceskip=0pt\relax}
\providecommand{\BIBentryALTinterwordstretchfactor}{4}
\providecommand{\BIBentryALTinterwordspacing}{\spaceskip=\fontdimen2\font plus
\BIBentryALTinterwordstretchfactor\fontdimen3\font minus
  \fontdimen4\font\relax}
\providecommand{\BIBforeignlanguage}[2]{{%
\expandafter\ifx\csname l@#1\endcsname\relax
\typeout{** WARNING: IEEEtran.bst: No hyphenation pattern has been}%
\typeout{** loaded for the language `#1'. Using the pattern for}%
\typeout{** the default language instead.}%
\else
\language=\csname l@#1\endcsname
\fi
#2}}
\providecommand{\BIBdecl}{\relax}
\BIBdecl

\bibitem{dzvonyar2018team}
D.~Dzvonyar, L.~Alperowitz, D.~Henze, and B.~Bruegge, ``Team composition in
  software engineering project courses,'' in \emph{2018 IEEE/ACM International
  Workshop on Software Engineering Education for Millennials (SEEM)}.\hskip 1em
  plus 0.5em minus 0.4em\relax IEEE, 2018, pp. 16--23.

\bibitem{dorodchi2019teaching}
M.~Dorodchi, E.~Al-Hossami, M.~Nagahisarchoghaei, R.~S. Diwadkar, and
  A.~Benedict, ``Teaching an undergraduate software engineering course using
  active learning and open source projects,'' in \emph{2019 IEEE Frontiers in
  Education Conference (FIE)}.\hskip 1em plus 0.5em minus 0.4em\relax IEEE,
  2019, pp. 1--5.

\bibitem{dumford2018online}
A.~D. Dumford and A.~L. Miller, ``Online learning in higher education:
  exploring advantages and disadvantages for engagement,'' \emph{Journal of
  Computing in Higher Education}, vol.~30, no.~3, pp. 452--465, 2018.

\bibitem{serrano2019technology}
D.~R. Serrano, M.~A. Dea-Ayuela, E.~Gonzalez-Burgos, A.~Serrano-Gil, and
  A.~Lalatsa, ``Technology-enhanced learning in higher education: How to
  enhance student engagement through blended learning,'' \emph{European Journal
  of Education}, vol.~54, no.~2, pp. 273--286, 2019.

\bibitem{ragusa2018sense}
A.~T. Ragusa and A.~Crampton, ``Sense of connection, identity and academic
  success in distance education: Sociologically exploring online learning
  environments,'' \emph{Rural Society}, vol.~27, no.~2, pp. 125--142, 2018.

\bibitem{seaman2018grade}
J.~E. Seaman, I.~E. Allen, and J.~Seaman, ``Grade increase: Tracking distance
  education in the united states.'' \emph{Babson Survey Research Group}, 2018.

\bibitem{allen2016online}
I.~E. Allen and J.~Seaman, \emph{Online Report Card: Tracking Online Education
  in the United States.}\hskip 1em plus 0.5em minus 0.4em\relax ERIC, 2016.

\bibitem{adedoyin2020covid}
O.~B. Adedoyin and E.~Soykan, ``Covid-19 pandemic and online learning: the
  challenges and opportunities,'' \emph{Interactive Learning Environments}, pp.
  1--13, 2020.

\bibitem{rodeghero2020empowering}
\BIBentryALTinterwordspacing
P.~Rodeghero and T.~Hernandez, ``Empowering and supporting remote software
  development team members through a culture of allyship,'' August 2020.
  [Online]. Available:
  \url{https://www.microsoft.com/en-us/research/publication/empowering-and-supporting-remote-software-development-team-members-through-a-culture-of-allyship-2/}
\BIBentrySTDinterwordspacing

\bibitem{ford2020tale}
D.~Ford, M.-A. Storey, T.~Zimmermann, C.~Bird, S.~Jaffe, C.~Maddila, J.~L.
  Butler, B.~Houck, and N.~Nagappan, ``A tale of two cities: Software
  developers working from home during the covid-19 pandemic,'' \emph{arXiv
  preprint arXiv:2008.11147}, 2020.

\bibitem{ardis2014software}
M.~Ardis, D.~Budgen, G.~Hislop, J.~Offutt, M.~Sebern, and W.~Visser, ``Software
  engineering 2014: curriculum guidelines for undergraduate degree programs in
  software engineering,'' \emph{joint effort of the ACM and the IEEE-Computer
  Society}, 2014.

\bibitem{shaw2005deciding}
M.~Shaw, J.~Herbsleb, I.~Ozkaya, and D.~Root, ``Deciding what to design:
  Closing a gap in software engineering education,'' in \emph{International
  Conference on Software Engineering}.\hskip 1em plus 0.5em minus 0.4em\relax
  Springer, 2005, pp. 28--58.

\bibitem{radermacher2014investigating}
A.~Radermacher, G.~Walia, and D.~Knudson, ``Investigating the skill gap between
  graduating students and industry expectations,'' in \emph{Companion
  Proceedings of the 36th international conference on software engineering},
  2014, pp. 291--300.

\bibitem{10.1145/3194779.3194780}
\BIBentryALTinterwordspacing
S.~Heggen and C.~Myers, ``Hiring millennial students as software engineers: A
  study in developing self-confidence and marketable skills,'' in
  \emph{Proceedings of the 2nd International Workshop on Software Engineering
  Education for Millennials}, ser. SEEM '18.\hskip 1em plus 0.5em minus
  0.4em\relax New York, NY, USA: Association for Computing Machinery, 2018, p.
  32–39. [Online]. Available:
  \url{https://doi-org.libproxy.clemson.edu/10.1145/3194779.3194780}
\BIBentrySTDinterwordspacing

\bibitem{stray2019slack}
V.~Stray, N.~B. Moe, and M.~Noroozi, ``Slack me if you can! using enterprise
  social networking tools in virtual agile teams,'' in \emph{2019 ACM/IEEE 14th
  International Conference on Global Software Engineering (ICGSE)}.\hskip 1em
  plus 0.5em minus 0.4em\relax IEEE, 2019, pp. 111--121.

\bibitem{li2016teaching}
Y.~Li, S.~Krusche, C.~Lescher, and B.~Bruegge, ``Teaching global software
  engineering by simulating a global project in the classroom,'' in
  \emph{Proceedings of the 47th ACM Technical Symposium on Computing Science
  Education}, 2016, pp. 187--192.

\bibitem{chatterjee2020software}
P.~Chatterjee, K.~Damevski, N.~A. Kraft, and L.~Pollock, ``Software-related
  slack chats with disentangled conversations,'' in \emph{IEEE International
  Working Conference on Mining Software Repositories}, 2020.

\bibitem{anewalt2017curriculum}
K.~Anewalt and J.~Polack, ``A curriculum model featuring oral communication
  instruction and practice,'' in \emph{Proceedings of the 2017 ACM SIGCSE
  Technical Symposium on Computer Science Education}, 2017, pp. 33--37.

\bibitem{saiedian2000requirements}
H.~Saiedian and R.~Dale, ``Requirements engineering: making the connection
  between the software developer and customer,'' \emph{Information and software
  technology}, vol.~42, no.~6, pp. 419--428, 2000.

\bibitem{jolak2020software}
R.~Jolak, M.~Savary-Leblanc, M.~Dalibor, A.~Wortmann, R.~Hebig, J.~Vincur,
  I.~Polasek, X.~Le~Pallec, S.~G{\'e}rard, and M.~R. Chaudron, ``Software
  engineering whispers: The effect of textual vs. graphical software design
  descriptions on software design communication,'' \emph{Empirical Software
  Engineering}, pp. 1--45, 2020.

\bibitem{jarboe1996procedures}
S.~Jarboe, ``Procedures for enhancing group decision making,''
  \emph{Communication and group decision making}, pp. 345--383, 1996.

\bibitem{kortum2017don}
F.~Kortum, J.~Kl{\"u}nder, and K.~Schneider, ``Don’t underestimate the human
  factors! exploring team communication effects,'' in \emph{International
  Conference on Product-Focused Software Process Improvement}.\hskip 1em plus
  0.5em minus 0.4em\relax Springer, 2017, pp. 457--469.

\bibitem{iacob2019exploring}
C.~Iacob and S.~Faily, ``Exploring the gap between the student expectations and
  the reality of teamwork in undergraduate software engineering group
  projects,'' \emph{Journal of Systems and Software}, vol. 157, p. 110393,
  2019.

\bibitem{blayone2017democratizing}
T.~J. Blayone, W.~Barber, M.~DiGiuseppe, E.~Childs \emph{et~al.},
  ``Democratizing digital learning: theorizing the fully online learning
  community model,'' \emph{International Journal of Educational Technology in
  Higher Education}, vol.~14, no.~1, pp. 1--16, 2017.

\bibitem{al2016investigating}
M.~Al-Emran, H.~M. Elsherif, and K.~Shaalan, ``Investigating attitudes towards
  the use of mobile learning in higher education,'' \emph{Computers in Human
  behavior}, vol.~56, pp. 93--102, 2016.

\bibitem{deFranco2017review}
J.~F. {DeFranco} and P.~A. {Laplante}, ``Review and analysis of software
  development team communication research,'' \emph{IEEE Transactions on
  Professional Communication}, vol.~60, no.~2, pp. 165--182, 2017.

\bibitem{storey2017how}
M.~{Storey}, A.~{Zagalsky}, F.~F. {Filho}, L.~{Singer}, and D.~M. {German},
  ``How social and communication channels shape and challenge a participatory
  culture in software development,'' \emph{IEEE Transactions on Software
  Engineering}, vol.~43, no.~2, pp. 185--204, 2017.

\bibitem{anders2016team}
A.~Anders, ``Team communication platforms and emergent social collaboration
  practices,'' \emph{International Journal of Business Communication}, vol.~53,
  no.~2, pp. 224--261, 2016.

\bibitem{seppala2016communication}
O.~Sepp{\"a}l{\"a}, T.~Auvinen, V.~Karavirta, A.~Vihavainen, and P.~Ihantola,
  ``What communication tools do students use in software projects and how do
  different tools suit different parts of project work?'' in \emph{2016
  IEEE/ACM 38th International Conference on Software Engineering Companion
  (ICSE-C)}.\hskip 1em plus 0.5em minus 0.4em\relax IEEE, 2016, pp. 432--435.

\bibitem{layng2016virtual}
J.~M. Layng, ``The virtual communication aspect: a critical review of virtual
  studies over the last 15 years,'' \emph{Journal of Literacy and Technology},
  vol.~17, no.~3, pp. 172--218, 2016.

\bibitem{handke2018medium}
L.~Handke, E.-M. Schulte, K.~Schneider, and S.~Kauffeld, ``The medium isn’t
  the message: Introducing a measure of adaptive virtual communication,''
  \emph{Cogent Arts \& Humanities}, vol.~5, no.~1, p. 1514953, 2018.

\bibitem{ford2019remotework}
\BIBentryALTinterwordspacing
D.~Ford, R.~Milewicz, and A.~Serebrenik, ``How remote work can foster a more
  inclusive environment for transgender developers,'' in \emph{Proceedings of
  the 2nd International Workshop on Gender Equality in Software Engineering},
  ser. GE '19.\hskip 1em plus 0.5em minus 0.4em\relax IEEE Press, 2019, p.
  9–12. [Online]. Available: \url{https://doi.org/10.1109/GE.2019.00011}
\BIBentrySTDinterwordspacing

\bibitem{jimenez2017working}
A.~Jimenez, D.~M. Boehe, V.~Taras, and D.~V. Caprar, ``Working across
  boundaries: Current and future perspectives on global virtual teams,''
  \emph{Journal of International Management}, vol.~23, no.~4, pp. 341--349,
  2017.

\bibitem{jarvenpaa1999communication}
S.~L. Jarvenpaa and D.~E. Leidner, ``Communication and trust in global virtual
  teams,'' \emph{Organization science}, vol.~10, no.~6, pp. 791--815, 1999.

\bibitem{rutowske2002}
A.~F. {Rutkowski}, D.~R. {Vogel}, M.~{Van Genuchten}, T.~M.~A. {Bemelmans}, and
  M.~{Favier}, ``e-collaboration: the reality of virtuality,'' \emph{IEEE
  Transactions on Professional Communication}, vol.~45, no.~4, pp. 219--230,
  2002.

\end{thebibliography}

\end{document}